\documentclass[sigconf]{acmart}
\AtBeginDocument{%
  }

\usepackage{acronym}
\usepackage{microtype}
\usepackage{graphicx}
\usepackage[inline]{enumitem}
\usepackage{csquotes}

\acrodef{ai}[AI]{Artificial Intelligence}
\acrodef{kb}[KB]{Knowledge Base}
\acrodef{llm}[LLM]{Large Language Model}
\acrodefplural{llm}[LLMs]{Large Language Models}
\acrodef{nlp}[NLP]{Natural Language Processing}
\acrodef{rag}[RAG]{Retrieval Augmented Generation}
\acrodef{sad}[SAD]{Software Architecture Documentation}
\acrodef{se}[SE]{Software Engineering}
\acrodef{tlr}[TLR]{Trace Link Recovery}

\setcopyright{acmlicensed}
\copyrightyear{2026}
\acmYear{2026}
\acmDOI{XXXXXXX.XXXXXXX}
\acmConference[ICSE-C '26]{International Conference on Software Engineering Companion}{April 12--18,
  2026}{Rio de Janeiro, Brazil}
\acmISBN{978-1-4503-XXXX-X/2018/06}

\begin{document}

\title{From Scattered to Structured: A Vision for Automating Architectural Knowledge Management}

\author{Jan Keim}
\email{jan.keim@kit.edu}
\orcid{0000-0002-8899-7081}
\affiliation{%
  \institution{Karlsruhe Institute of Technology (KIT)}
  \country{Germany}
}

\author{Angelika Kaplan}
\email{angelika.kaplan@kit.edu}
\orcid{0009-0009-9101-5833}
\affiliation{%
  \institution{Karlsruhe Institute of Technology (KIT)}
  \country{Germany}
}

\renewcommand{\shortauthors}{Keim \& Kaplan}

\begin{abstract}
Software architecture is inherently knowledge-centric.
The architectural knowledge is distributed across heterogeneous software artifacts such as requirements documents, design diagrams, code, and documentation, making it difficult for developers to access and utilize this knowledge effectively.
Moreover, as systems evolve, inconsistencies frequently emerge between these artifacts, leading to architectural erosion and impeding maintenance activities.
We envision an automated pipeline that systematically extracts architectural knowledge from diverse artifacts, links them, identifies and resolves inconsistencies, and consolidates this knowledge into a structured knowledge base.
This knowledge base enables critical activities such as architecture conformance checking and change impact analysis, while supporting natural language question-answering to improve access to architectural knowledge.
To realize this vision, we plan to develop specialized extractors for different artifact types, design a unified knowledge representation schema, implement consistency checking mechanisms, and integrate retrieval-augmented generation techniques for conversational knowledge access.

\end{abstract}




\maketitle

\section{Introduction}
\label{sec:intro}


Software architecture constitutes a fundamental pillar of software engineering \cite{medvidovic2010}, embodying the high-level structure and design decisions that shape a system's quality attributes and long-term maintainability \cite{kruchten2004,keim2022taxonomy}.
However, architectural knowledge is inherently complex, drawing from both theoretical principles and practical experience accumulated throughout a system's lifecycle. \cite{jansenbosch2005,kruchten2006,farenhorst2009}
This knowledge manifests across different artifacts: requirements documents capture needs and constraints, design diagrams visualize system structures and component interactions, code comments embed implementation rationale, and documentation describes architectural decisions and their justifications.

The distribution of architectural knowledge across these heterogeneous artifacts presents significant challenges \cite{farenhorst2009,keim2019towards}.
While some knowledge is explicitly documented, substantial portions remain implicit, requiring inference from context or existing only in developers' minds \cite{kruchten2006}.
Additionally, external sources contribute essential knowledge through established architectural styles, design patterns, and industry best practices.
Accessing and effectively utilizing this scattered knowledge requires considerable effort from developers and architects, who must spend significant time and resources on comprehension \cite{xia2018}, particularly in large-scale systems where documentation may span thousands of pages and codebases may contain millions of lines.

Compounding these accessibility challenges, inconsistencies frequently arise across artifacts as systems evolve \cite{keim2023icsa, wohlrab2019}.
Documentation may fall out of sync with implementation, design diagrams may not reflect recent architectural changes, and code comments may describe outdated design decisions.
Such inconsistencies breed confusion, impede maintenance activities, and can lead to architectural erosion, in which the implemented system gradually deviates from its intended design.

\begin{figure*}
\begin{center}
    \includegraphics[width=0.8\linewidth]{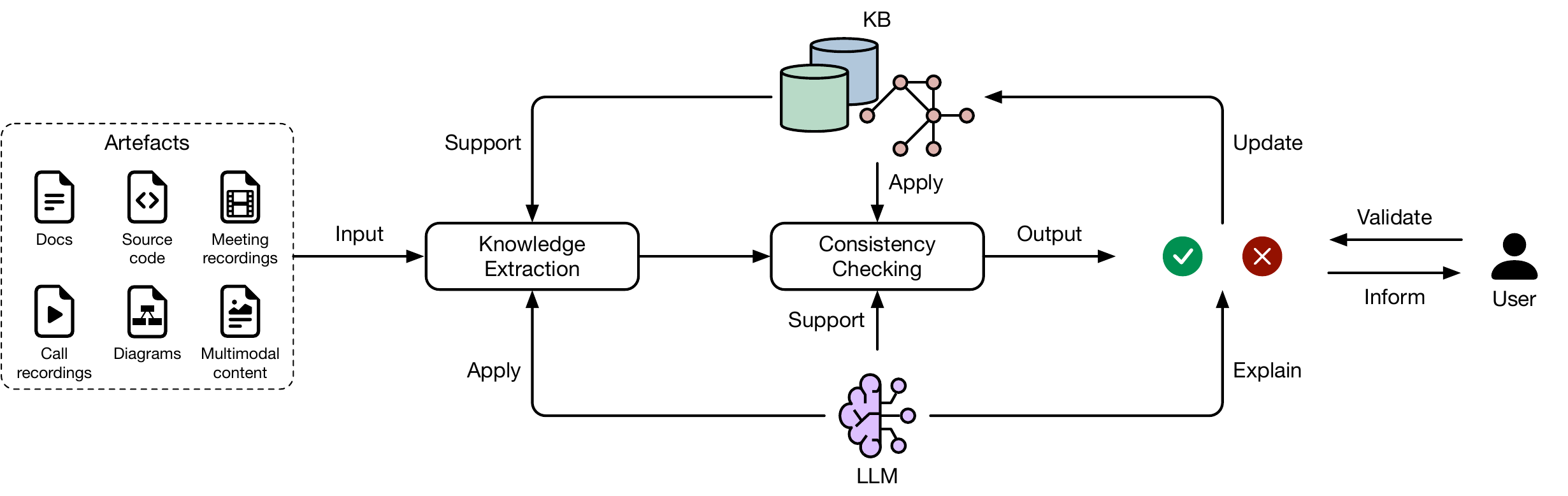}
\caption{Artifacts knowledge extraction and consistency checking pipeline, by Keim et al. \cite{Keim2025_TowardsKM}}
\Description{A diagram showing the envisioned pipeline for extracting architectural knowledge from diverse software artifacts, checking for inconsistencies, and consolidating this knowledge into a structured knowledge base that supports question-answering and continuous monitoring.}
\label{fig:pipeline}
\end{center}
\end{figure*}

To address these challenges, we envision an approach (see \autoref{fig:pipeline}) that systematically extracts architectural knowledge from diverse software artifacts, identifies and resolves inconsistencies, and consolidates this knowledge into a structured knowledge base \cite{Keim2025_TowardsKM}.
The resulting knowledge base can support critical activities, including architecture conformance checking, change impact analysis, and informed decision-making during architectural evolution.
In addition to software artifacts like code and documentation, our long-term vision includes considering other sources, such as meeting minutes and recordings.
Furthermore, by building upon structured knowledge representations, we can enable question-answering systems that allow developers and architects to query architectural knowledge through natural language, analogous to systems for scientific knowledge retrieval such as HubLink \cite{hublink}.

We also envision this system operating as an intelligent agent that continuously monitors software artifacts, automatically extracts knowledge, checks consistency across sources, and updates the knowledge base in real-time.
When the automated analysis encounters ambiguities or conflicts that require domain expertise, the agent can proactively engage architects or developers through targeted questions, ensuring that human judgment informs critical decisions while minimizing the burden on development teams.
Such capabilities promise to democratize access to architectural knowledge and support more informed software engineering practices throughout the development lifecycle.
Additionally, this might encourage better documentation practices, as developers recognize the tangible benefits of maintaining accurate and consistent architectural knowledge.

The remainder of this paper is structured as follows:
\autoref{sec:rw} reviews the state-of-the-art and existing related work.
\autoref{sec:steps} outlines our planned steps to develop the automated knowledge extraction pipeline, including the agent-based monitoring approach, extraction framework, and knowledge base design. 
\autoref{sec:challenges} discusses anticipated challenges and open issues.

\section{State-of-the-Art and Related Work}
\label{sec:rw}

Managing and utilizing knowledge across various software artifacts is a fundamental challenge in software engineering.
Throughout the development and maintenance lifecycle, substantial knowledge is embedded in diverse artifacts such as requirements documents, architectural models, source code, and test cases.
Effectively leveraging this distributed knowledge and maintaining consistency across artifacts requires sophisticated techniques and tools.
In the following subsections, we review key research areas that form the foundation for automated architectural knowledge management: \ac{tlr} for linking artifacts, consistency checking for detecting conflicts, knowledge representation approaches, the emerging role of large language models, and question-answering systems for knowledge access.

\subsection{Traceability and Knowledge Integration}

Trace links and their recovery (\acs{tlr}) serve as a primary mechanism for identifying overlapping knowledge and integrating information scattered across different artifacts.
These links support critical software engineering activities, including maintenance \cite{clelandHuang2012}, bug localization \cite{rath18}, change impact analysis \cite{falessi20}, and consistency checking \cite{stulova2020,keim2023icsa}.

Researchers have developed diverse techniques for \ac{tlr}, including information retrieval approaches \cite{hayes2006,hey2021_ftlr} and machine learning techniques \cite{guo2017,mills2018}.
More recently, \acp{llm} have emerged as a powerful tool for traceability \cite{lin_traceability_2021,rodriguez2023,hey2025,fuchss2025}.

\subsection{Consistency and Conformance Checking}

Established trace links enable inconsistency detection across artifacts, as exemplified by Keim et al.~\cite{keim2023icsa}.
Similarly, Bucaioni et al.~\cite{bucaioni2024} employ mappings to verify architectural conformance.
Ensuring that implementations reflect their specifications and that knowledge remains consistent across all artifacts is crucial for software quality.
Consequently, extensive research addresses conformance checking to verify that code and system behavior align with specifications, architectures, and other normative artifacts (e.g., \cite{pruijt2014,schroeder18,ciraci2012,li1998}).

The model-driven development community has particularly emphasized consistency across different models and system views, especially in complex domains such as cyber-physical systems \cite{reussner2023}.
Inconsistent models and views impede development, yet manually maintaining model consistency is challenging when numerous models coexist.
Research in this area includes bidirectional model transformations \cite{stevens2008}, change-propagation mechanisms using consistency-preservation rules \cite{kramer2015}, and approaches that integrate multiple models into unified views \cite{atkinson2008,burger2013,meier2019}.

\subsection{Knowledge Bases and Knowledge Management}

Knowledge bases provide structured approaches to support software engineering activities \cite{lowry1993,ding2014}.
Knowledge-based development facilitates systematic software evolution by explicitly capturing and utilizing domain and system knowledge.
Ontologies offer a common representation mechanism for knowledge storage, leading to various ontology-based software engineering approaches \cite{dillon2008}.
Knowledge graphs extend these concepts by representing entities and their relationships as graphs, providing a powerful foundation for question-answering systems (cf.~\autoref{sec:qa}) in which structured knowledge can be queried and combined with generative capabilities.
Additionally, search-based software engineering techniques enable knowledge retrieval through queries \cite{harman2012}.

\subsection{Large Language Models in Software Engineering}

\Acp{llm} have recently transformed software engineering practices through their capacity to process code and support diverse development tasks \cite{jin2024,he2025}.
Notably, \acp{llm} can provide architectural knowledge and respond to developer queries \cite{soliman2025}, making them a promising technology for bridging knowledge across different artifacts and supporting automated knowledge management throughout the software lifecycle.

\subsection{Question-Answering Systems}
\label{sec:qa}
Question-answering (QA) systems enable users to retrieve specific information via natural language queries without requiring familiarity with the underlying data structures, making them valuable in software engineering contexts where developers need quick access to architectural knowledge and design decisions from large repositories.
Knowledge graphs provide structured foundations for QA systems, while retrieval-augmented generation (RAG) approaches combine information retrieval with generative capabilities to produce accurate answers.
Systems like HubLink, introduced by Kaplan et al.~\cite{hublink}, demonstrate effective QA retrieval over knowledge graphs by combining graph-based retrieval with language-model capabilities.
Adapting such approaches to software engineering knowledge bases could bridge the gap between scattered documentation and developers' information needs, supporting informed decision-making throughout the software lifecycle.

\section{Planned Steps}
\label{sec:steps}

To realize our vision, we will conduct a comprehensive literature review to further establish theoretical foundations, identify state-of-the-art approaches to architectural knowledge management and extraction from diverse artifacts, and derive a taxonomy of architectural knowledge types and an understanding of current limitations.
To realize the vision of an automated pipeline for architectural knowledge extraction and consolidation, we propose the following systematic approach, where we will empirically evaluate each step to ensure effectiveness and practicality:

\subsection{Knowledge Extraction Framework}
We will develop specialized extractors for different artifact types, each tailored to the unique characteristics of its source.
Textual artifacts (e.g., requirements and documentation) will be processed using \ac{nlp} techniques and \acp{llm} to identify architectural elements, decisions, and rationale. 
Static analysis and program comprehension techniques will extract structural and behavioral knowledge from code artifacts.
Visual artifacts such as design diagrams require parsing mechanisms to capture component relationships and architectural patterns.
The expected outcome is a modular extraction framework capable of processing heterogeneous software artifacts and producing structured knowledge representations.
To address inevitable gaps in automated extraction, such as missed key architectural terms, incorrectly inferred links, or project-specific domain knowledge, the framework will support Human-AI collaboration:
Manual knowledge augmentation through user inputs (e.g., structured JSON documents) will allow users to provide missing information, correct extraction errors, and refine the knowledge base with project-specific context.

\subsection{Knowledge Base Schema Design}
We will design a unified schema for the architectural \ac{kb} that can accommodate knowledge from diverse sources while maintaining semantic relationships and traceability links through \ac{tlr}.
This schema has to support queries for architecture conformance checking, change impact analysis, and natural language question-answering.
To link different artifacts, we will use traceability techniques from previous work \cite{keim2021ecsa,keim2024icse,fuchss2025}.

\subsection{Consistency Checking and Resolution}
We will reuse existing and develop new automated mechanisms to identify inconsistencies between knowledge extracted from different artifacts.
This includes detecting contradictions in architectural decisions, mismatches between documented and implemented structures, and outdated information.
For example, in previous work, Keim et al.~\cite{keim2023icsa} demonstrated the use of traceability links for inconsistency detection, which we will adapt and extend for our multi-artifact context.
We will explore rule-based, learning-based, and \ac{llm}-based approaches to inconsistency detection and develop strategies for semi-automated resolution that involve human architects in critical decisions.

\subsection{Agent-Based Continuous Monitoring}
To operationalize the knowledge management pipeline, we will develop an \ac{llm} agent that continuously monitors software artifacts and orchestrates the knowledge extraction and maintenance workflow.
The agent will detect changes in artifacts through file system monitoring and version control integration, automatically triggering \ac{tlr} for modified or newly created artifacts to maintain up-to-date traceability across the system.
When inconsistencies are detected through automated consistency checking, the agent will analyze their severity and context, either resolving minor issues autonomously or escalating critical conflicts to human architects with targeted clarification requests.
For ambiguous architectural decisions or conflicting information that requires domain expertise, the agent will formulate specific questions to architects or developers, presenting relevant context and suggesting potential resolutions.
Before executing actions that significantly impact the knowledge base, such as removing outdated information or merging conflicting architectural decisions, the agent will seek explicit confirmation from responsible stakeholders.
The expected outcome is an autonomous system that maintains knowledge base integrity with minimal human intervention while ensuring that critical decisions remain under human control.

\subsection{Question-Answering System Integration}
Building upon the structured \ac{kb}, we will develop a natural language question-answering interface using \ac{rag} techniques.
This will enable developers and architects to query architectural knowledge conversationally, similar to systems like HubLink~\cite{hublink} for scientific knowledge (see \autoref{sec:qa}).
We will evaluate the system's ability to answer common architectural queries accurately and efficiently.
The expected outcome is a prototype system that democratizes access to architectural knowledge across development teams.

\section{Challenges}
\label{sec:challenges}

The envisioned approach presents several significant challenges that have to be addressed to ensure its effectiveness and reliability.

\textbf{Heterogeneous artifact formats and multimodal data.}
Architectural knowledge manifests across diverse artifact types, structured sources such as code and formal documentation, partially structured sources such as design diagrams and models, and unstructured sources, including informal documentation, meeting transcripts, sketches, and informal notes.
Processing the knowledge from these heterogeneous and multimodal sources requires techniques that can handle varying levels of formality and different representation formats without extensive manual preprocessing.

\textbf{Scalability and context limitations.}
Large-scale systems may contain millions of lines of code and thousands of pages of documentation, which can challenge both computational resources and the context windows of \acp{llm} used for knowledge extraction.
The pipeline must efficiently process large artifact collections while maintaining accuracy, and determine optimal strategies for partitioning and processing artifacts that exceed model context limits.
Therefore, techniques and approaches for effectively managing memory and context are essential.

\textbf{Explainability and trust.}
\Ac{llm}-based extraction approaches are susceptible to hallucinations, generating plausible but factually incorrect information.
For architectural knowledge management, such errors can have serious consequences if they propagate into the knowledge base and influence critical decisions.
Ensuring the explainability of extracted knowledge and building trust through verification mechanisms, like combining \acp{llm} with formal reasoning tools and ontologies, is essential for practical adoption.

\textbf{Distinguishing extraction errors from genuine inconsistencies.}
When the automated approach detects inconsistencies, it has to distinguish true conflicts in the artifacts (e.g., outdated documentation) from errors introduced during the extraction process.
This distinction is crucial for determining appropriate resolution strategies and avoiding false alarms that burden developers.
Furthermore, there may be planned inconsistencies, such as architectural changes that are not yet implemented, which the system must recognize and handle appropriately.

\textbf{Human-AI collaboration and feedback integration.}
The agent-based approach has to strike an appropriate balance between automation and human oversight.
Excessive automation without sufficient human validation may introduce errors, while excessive human intervention negates efficiency benefits.
Designing effective interaction patterns for clarification requests, confirmation workflows, and feedback integration requires careful consideration of developer workflows and cognitive load.

\textbf{Contextual learning and project-specific adaptation.}
Each software project exhibits unique architectural patterns, naming conventions, and documentation practices.
The knowledge extraction pipeline has to adapt to project-specific contexts using techniques such as few-shot learning and incremental refinement guided by developer feedback, rather than relying solely on general-purpose models trained on external data.

\textbf{Knowledge base performance and maintenance.}
As the knowledge base grows with continuous artifact monitoring and extraction, maintaining query efficiency becomes critical to support real-time architecture conformance checking and responsive question-answering systems.
The pipeline has to implement effective update management strategies to handle incremental knowledge additions while ensuring consistency, managing versioning, and avoiding performance degradation.
Furthermore, when processing sensitive artifacts such as meeting minutes or voice recordings, data privacy considerations require appropriate access controls and anonymization techniques.

\textbf{Temporal validity and system evolution.}
Software systems evolve continuously, and some evolutionary steps are dramatic, such as migrating to entirely different architecture styles and technology stacks.
Despite these transformations, the system remains conceptually the same entity, yet its architectural form changes fundamentally.
This creates a significant challenge for knowledge management: architectural knowledge that is valid at one time may become obsolete or incorrect at another time, yet it represents important historical context.
The approach must address whether to discard outdated information, thereby losing valuable historical insights and evolution rationale, or to maintain a temporal knowledge base with versioning and time-aware query capabilities.
This requires sophisticated mechanisms to track changes over time and ensure that queries return contextually relevant information based on the system's state at specific points in its evolution.

\textbf{Security and Access Management.}
When processing sensitive artifacts such as meeting minutes or voice recordings, data privacy considerations require appropriate access controls and anonymization techniques.
Moreover, ensuring that the knowledge extraction and storage processes comply with organizational security policies is essential to protect sensitive architectural information from unauthorized access or breaches.

\begin{acks}
This work was funded by the Topic Engineering Secure Systems of the Helmholtz Association (HGF) and by the Deutsche Forschungsgemeinschaft (DFG, German Research Foundation) under the National Research Data Infrastructure -- NFDI 52/1 -- project number 501930651, NFDIxCS.
It was also supported by funding from the pilot program Core Informatics at KIT (KiKIT) of the Helmholtz Association (HGF) and by KASTEL Security Research Labs.
\end{acks}

\bibliographystyle{ACM-Reference-Format}
\bibliography{paper}

\end{document}